\documentclass{elsart}

\usepackage{epsfig}

\usepackage{amssymb}

\begin{document}
\newcommand{\be}{\begin{equation}}
\newcommand{\ee}{\end{equation}}

\begin{frontmatter}



\title{Peak finding through Scan Statistics}


\author[lnf]{Francesco Terranova\corauthref{cor}} 
\corauth[cor]{Corresponding author.}
\address[lnf]{Laboratori
Nazionali di Frascati dell'INFN, Frascati (Roma), Italy}

\begin{abstract}
We discuss the conditions under which Scan Statistics can be
fruitfully implemented to signal a departure from the underlying
probability model that describes the experimental data. It is shown
that local perturbations (``bumps'' or ``excesses'' of events) are
better dealt within this framework and, in general, tests based on
these statistics provide a powerful and unbiased alternative to the
traditional techniques related with the $\chi^2$ and Kolmogorov
distributions. Approximate formulas for the computation of Scan
Statistics in the range of interest for high energy and nuclear
physics applications are also presented. 
\end{abstract}

\end{frontmatter}

\section{Introduction}
\label{sec:introduction}
In the last two decades, the properties of ``scan statistics''
\cite{glaz_naus,glaz_balakrishnan} have been extensively investigated
and this subject still represents a rapidly developing branch of
applied probability. Nowadays, scan statistics are used in several
areas of science and engineering to analyze the occurrence of cluster
of events and to assess their statistical significance. Applications
range from control theory to molecular biology while, at present,
their use in physics is mainly limited to the analysis of time series,
especially in x and $\gamma$-ray astronomy \cite{orford}. In fact, a
common problem in nuclear and particle physics is to determine whether
an observed cluster of events has occurred by chance or if it signals
a departure from the underlying probability model (``null
hypothesis'') for the data. Peaks or local excesses of events can
appear during energy scans at the colliders, in the distributions of
kinematical variables as invariant masses or four-momentum transfer,
in the interpretation of Dalitz plots, etc.  The traditional test
statistics that are employed to challenge the null hypothesis after a
data taking can be divided into two broad families. Binned
goodness-of-fit tests are connected to the $\chi^2$ distribution. If
the region where the excess is expected is known {\it a priori}
because it has already been observed by another experiment, i.e. a
confirmatory experiment against a claim of discovery has been
performed, the level of agreement between the null hypothesis and the
data can be evaluated in a straightforward manner through a Pearson
$\chi^2$ test~\cite{eadie}. For one-dimensional distributions of the
variable $x$ ($x\in \left[\mathcal{A},\mathcal{B}\right]$), let the
number of observations in the candidate signal region
$\left[a,b\right]$ be $k_{ab}$. The best estimate of the background in
the region $\left[a,b\right]$ is given by

\be 
\hat{B}_{ab} = \int_{a}^{b} B(x,\hat{\underline{\theta}}) \d x
\label{equ:b_ab}
\ee

\noindent where $\underline{\theta}$ is the set of parameters
describing the background distribution and $\hat{\underline{\theta}}$
is its best estimate based on events outside the interval
$\left[a,b\right]$, so that
$\mathrm{cov}(k_{ab},\hat{B}_{ab})=0$. Hence, the test variable can be
defined as:

\be
T_{ab} \equiv   \frac{(k_{ab}-\hat{B}_{ab})^2}{V(k_{ab}-\hat{B}_{ab})}
\ee

\noindent $V(k_{ab}-\hat{B}_{ab})$ being the variance of
$k_{ab}-\hat{B}_{ab}$. Under the null hypothesis
$V(k_{ab})=\hat{B}_{ab}$ and
$V(k_{ab}-\hat{B}_{ab})=\hat{B}_{ab}+\hat{\sigma}^2_{ab}$, i.e.  the
variance is the quadratic sum of the estimated background rate in the
signal region and its error. Finally, if the error on the background
is negligible

\be
T_{ab} \simeq
\frac{(k_{ab}-\hat{B}_{ab})^2}{\hat{B}_{ab}}
\label{equ:conf_experiment}
\ee

\noindent and, in the asymptotic limit, the test variable behaves as
$\chi^2(1)$.  Clearly, if the bin number $N_\mathrm{bin}$ and the bin
size $w\equiv(\mathcal{B}-\mathcal{A})/N_\mathrm{bin}$ are specified
in advance but no information on the position, size and width of the
signal are available, the corresponding test is given by

\be
T\equiv \sum_{i=1}^{N_\mathrm{bin}} \frac{(k_i-b_i)^2}{b_i}
\label{equ:test_chi2}
\ee

\noindent which behaves as $\chi^2(N)$ in the asymptotic limit.  The
power\footnote{The power of an hypothesis test against a specific
alternative hypothesis is the chance that the test correctly rejects
the null hypothesis when that alternative hypothesis is true; that is,
the power is 100\% minus the chance of a Type~II error when that
alternative hypothesis is true.} of this test depends on the position
and width of the signal compared to the binning since a cluster shared
among several bins becomes harder to be detected while clusters
appearing in too large bins are swamped by background. If the size and
position of the bin is chosen {\it after} the inspection of the data,
the power of the test is increased but the estimate of the
$p$-value\footnote{For a null hypothesis $H_0$ and a test statistic
$T$ (e.g. the one of Eq.~(\ref{equ:test_chi2})~)  we define $g(T|H_0)$ as
the p.d.f. of $T$ in the occurrence of $H_0$. The $p$-value is defined
as the probability to find $T$ in the region of equal or less
compatibility with $H_0$ than the level of compatibility observed with
the actual data~\cite{pdg}. For example, if $T$ is defined such
that large values correspond to poor agreement with the hypothesis,
then the $p$-value will be
$$
\int_{T_{obs}}^{+\infty} g(T | H_0) dT
$$ 
$T_{obs}$ being the value of the test statistic obtained in the
actual experiment.}  of the null hypothesis inferred from $\chi^2(N)$
is unreliable; hence, the significance level of the hypothesis test is
biased.  The binning problem can be overcome employing a second family
of test statistics connected to the Kolmogorov distribution. The most
common test is the Kolmogorov-Smirnov test \cite{eadie} which
corresponds to the largest distance between the cumulative
distribution of the data and the one of the null hypothesis. This
distance has a characteristic distribution that can be computed
analytically and, hence, provides the $p$-value of the null hypothesis
for a given data taking. These tests are well-suited to detect global
distortions of the $x$ distribution but have limited power for strong
local deviation from the null hypothesis (cluster of
events). Moreover, the power depends significantly on the position of
the signal peak.

The procedure used to seek for event clusters {\it after} the data
taking suggests a possible alternative to those methods, but which
does not suffer from the problems of choosing the region {\it a
posteriori}. The search is performed scanning the
$\left[\mathcal{A},\mathcal{B}\right]$ interval to identify the region
where an anomalous accumulation of events appears. Given $N$ events
distributed along the $\left[\mathcal{A},\mathcal{B}\right]$ range, we
call $S(w)$ the largest number of events in a window of fixed length
$w$. If the distribution of $S(w)$ is known, it will be possible to
compute the probability $Prob(S(w) \geq k)$ for the null hypothesis to
produce a cluster $S(w)$ greater or equal than the one actually
observed. Hence, the $p$-value of the null hypothesis can be
assessed. In this context, an {\it a priori} binning similar to the
one of the Pearson $\chi^2$ test is no more needed. Moreover, the test
statistics $S(w)$ (``scan statistics'') is not affected by the
drawbacks of the Kolmogorov-Smirnov (K-S) tests (see
Sec.~\ref{sec:comparison}). Sometimes this approach is followed, at
least qualitatively, in literature. For instance, the OPAL~\cite{OPAL}
data accumulated at LEP during the high energy run beyond the $Z^0$
resonance (LEP2) were used to falsify the ALEPH~\cite{ALEPH} claim of
a peak in the dijet invariant mass $M$ of the $e^+e^-
\rightarrow$~four jets final state at $M\simeq 105$~GeV. Clearly, this
refutation was carried out using the test statistics of
Eq.~(\ref{equ:conf_experiment}).  Moreover, to test for a peak in the
dijet mass sum distribution for arbitrary mass $M$ and independent of
histogram binning, the positions of the mass windows were scanned over
the full range of $M$. However, no quantitative statement was drawn
due to the strong correlations of the contents of nearby bins. In
fact, accounting for this correlation is possible once the properties
of the scan statistics $S(w)$ are determined.  In
Sec.~\ref{sec:scanstat} these properties are discussed and the
formulas to compute $Prob(S(w) \geq k)$ are presented.  The power and
significance of the test statistics $S(w)$ is computed in
Sec.~\ref{sec:comparison} and compared with the Pearson $\chi^2$ and
K-S approach for one-dimensional distributions. Extensions of the
tests based on $S(w)$ and further applications in particle physics
data analyses are discussed in Sec.~\ref{sec:extensions}.

\section{Scan statistics}
\label{sec:scanstat}

Let us consider an interval $\left[\mathcal{A},\mathcal{B}\right]$ of
a continuous variable $x$ and a Poisson process (``background'') whose
mean value per unit interval is denoted with $\lambda$.  Hence, the
probability of finding $Y_x(w)$ events in an interval
$\left[x,x+w\right]$ is \be Prob(Y_x(w)=k) \ = \ e^{-\lambda w}
\frac{(\lambda w)^k}{k!} \ ; \ \ \ \ k=0,1,2,\ldots
\label{equ:poisson}
\ee

\noindent The number of events in any disjoint non-overlapping
intervals are independently distributed. We call ``scan statistic''
(SS) the largest number of events to be found in any subinterval of
$\left[\mathcal{A},\mathcal{B}\right]$ of length $w$~\footnote{The
case of non-uniform background can be dealt with by allowing for a
window of variable width $w(x)$ that always contains
$w/(\mathcal{B}-\mathcal{A})$ percent of the expected events under the
null hypothesis~\cite{weinstock}.}, i.e.

\be
S(w) \equiv \max_{\mathcal{A}\leq x \leq \mathcal{B}-w} \left\{ Y_x(w) \right\} 
\ee

\noindent The probability that the number of events in a scanning window
never reaches $k$ will be denoted, following~\cite{glaz_naus}, as

\be
Q^*(k,\lambda \Delta ,w/\Delta) \equiv 1-Prob(S(w)\geq k) 
\label{equ:qstar}
\ee

\noindent where $\Delta\equiv \mathcal{B}-\mathcal{A}$ and the suffix
``*'' indicates that unconditional probabilities are considered,
i.e. that the overall number of events $N$ in the interval
$\left[\mathcal{A},\mathcal{B}\right]$ is not fixed but it fluctuates
according to Eq.~(\ref{equ:poisson}) with $w=\Delta$.  The exact form
of Eq.~(\ref{equ:qstar}) can be expressed in terms of the sum of
products of two determinants~\cite{huntington_naus}.  The summation is
over the set $V$ of all the partitions of $N$ into $2H+1$ non-negative
integers $m_i$ satisfying $m_{i}+m_{i+1}<k$ for $i=1,\ldots,2H$, where
$H$ is the largest integer in $\Delta/w$.  The determinants are
computed starting from the $(H+1) \times (H+1)$ matrix $\{h\}_{ij}$
and the $H \times H$ matrix $\{v\}_{ij}$ whose entries are:
\begin{eqnarray*}
h_{ij} & = & \;\;\;\sum_{s=2j-1}^{2i-1}\!\!m_{s}-(i-j)k\;\;\;\; 
1 \leq j \leq i \leq H+1
\\
       & = & -\sum_{s=2i}^{2j-2}\!\!m_{s}+(j-i)k\;\;\;\; 1\leq i<j\leq H+1
\\
v_{ij} & = & \;\;\;\sum_{s=2j}^{2i}m_{s}-(i-j)k\;\;\;\;1 \leq j \leq i \leq H
\\
       & = & -\sum_{s=2i+1}^{2j-1}\!\!m_{s}+(j-i)k\;\;\;\;1\leq i<j\leq H
\end{eqnarray*}

Using these definitions for $V$, $h_{ij}$ and $v_{ij}$, we have
for $k\geq2$ and $w<\Delta$:

\be
Q^*(k,\lambda \Delta ,w/\Delta)= \sum_{V} R^* \ \mathrm{det}|1/h_{ij}!| \ 
\mathrm{det}|1/v_{ij}!|
\label{equ:exact}
\ee

In formula~(\ref{equ:exact}) 
\be
R^*=N!\ d^{M} \ (\frac{w}{\Delta}-d)^{N-M} \ p(N,\lambda \Delta)
\ee
\be
M=\sum_{j=0}^{H}m_{2j+1}
\ee
\noindent being $d\equiv1-wH/\Delta$ and $p(N,\lambda \Delta)$ is the Poisson
probability of having $N$ events from an average rate $\lambda
\Delta$. 

\noindent A very useful approximation of Eq.~(\ref{equ:exact})
has been derived by Naus in 1982~\cite{naus82}, based on the exact
values of the probabilities $Q_2 \equiv Q^*(k,2\psi ,1/2)$ and $Q_3
\equiv Q^*(k,3\psi ,1/3)$)~\footnote{For later convenience we define
$\psi\equiv \lambda w$ and $L=\Delta/w$.}.  It can be shown that

\begin{equation}
Q^{*}(k;\psi L,1/L) \simeq Q^{*}_2\left[Q^{*}_3/Q^{*}_2\right]^{L-2}
\label{equ:naus}
\end{equation}

\noindent where

\begin{eqnarray}
Q^{*}_2 & = & \left[F(k-1,\psi)\right]^{2}-
\left(k-1\right)p(k,\psi)p(k-2,\psi) \nonumber
\\
& & -\left(k-1-\psi\right)p(k,\psi)F(k-3,\psi)
\label{equ:Q2_naus} \\
Q^{*}_3 & = & \left[F(k-1,\psi)\right]^{3}-A_{1}+A_{2}+A_{3}-A_{4} 
\label{equ:Q3_naus} 
\end{eqnarray}

\noindent and

\begin{eqnarray*}
A_{1} & = & 2\ p(k,\psi)F(k-1,\psi)\left\{\left(k-1\right)F(k-2,\psi)-\psi F(k-3,\psi)\right\}
\\
A_{2} & = & 0.5 \ \left[p(k,\psi)\right]^{2}\left\{\left(k-1\right)\left(k-2\right)F(k-3,\psi) \right. 
\\ 
& & \left.
-2\left(k-2\right)\psi F(k-4,\psi)+\psi^{2}F(k-5,\psi)
\right\}
\\
A_{3} & = & \sum_{r=1}^{k-1}p(2k-r,\psi)\left[F(r-1,\psi)\right]^{2}
\\
A_{4} & = & \sum_{r=2}^{k-1}p(2k-r,\psi)p(r,\psi)
\left\{\left(r-1\right)F(r-2,\psi)-\psi F(r-3,\psi)\right\}
\end{eqnarray*}

\noindent
In the above formulas $F(k,\psi)$ denotes the cumulative distribution

\be
F(k,\psi) = \sum_{i=0}^k p(i,\psi) \ \ ; \ \ p(i,\psi)=e^{-\psi} 
\frac{\psi^i}{i!}
\ee

\noindent and $ F(k,\psi)=0$ for $k<0$. For large values of $\Delta/w$
an even simpler approximation due to Alm~\cite{alm} can be implemented:

\begin{eqnarray}
Q^*(k,\lambda \Delta, w/\Delta) \simeq & \nonumber \\ & \hspace{-1.5cm}
F(k-1,\lambda w) 
\exp\left\{-\frac{k-w\lambda}{k}\lambda(\Delta-w)\ p(k-1,\lambda w)\right\}
\label{equ:alm}
\end{eqnarray}

\noindent Eq.~(\ref{equ:alm}) is often used in astrophysics applications
and in many time series problems but it is of limited use in the
present case where the condition $w/\Delta \ll 1$ is rarely fulfilled.
In the following, the test statistics based on SS will be studied
relying on the approximation (\ref{equ:naus}). For a systematic
comparison of the various approximations of Eq.~(\ref{equ:exact}) we
refer to~\cite{glaz_naus}.

\section{Power and significance for one-dimensional distributions}
\label{sec:comparison}

Sec.~\ref{sec:scanstat} dealt with the distribution of the scan
statistics under the null hypothesis. The class of alternative
hypotheses considered hereafter describe a local perturbation of the
uniform distribution which leads to the appearance of a ``excess'' of
events. Long-range distortions like anomalous angular distributions
are better dealt with global K-S tests and will not be further
considered here.  The alternative functions are Poisson processes of
mean $S$. The signal events are spread along
$\left[\mathcal{A},\mathcal{B}\right]$ according to a normal
distribution of mean $x_S$ and sigma $\sigma_S$~\footnote{This is the
case, for instance, of a narrow resonance whose intrinsic width is
smaller than the detector resolution. For resonances broader than the
instrumental precision a relativistic Breit-Wigner or a Jacobian-peak
would be more appropriate. However, for the present purposes the
details of the alternative function are not critical.}. In the
following we reject the null hypothesis if its $p$-value is smaller
than 5\%. The actual significance of the test statistics, i.e. the
number of experiments where the null hypothesis was rejected albeit
true, has been computed by Monte Carlo experimentation. Similarly the
rate of Type~II errors was computed to estimate the power of the test.
In general, some prior assumptions are made before the inspection of a
distribution.  The domain $\left[\mathcal{A},\mathcal{B}\right]$ of
the variable $x$ accessible to the experiment depends on the
particular apparatus and, in most of the cases, it is selected {\it a
priori}; so it is not expected to be a source of biases. In fact, in
many applications the final distribution of $x$ is the result of
sequential cuts on other kinematic variables which can severely bias
the sample~\cite{blind_analysis}. For narrow resonances, whose width
does not exceed the (known) instrumental resolution of the detector,
the scanning window $w$ of $S(w)$ can be fixed {\it a priori}. In
particular, for a gaussian perturbation of variance $\sigma_S^2$ a
nearly optimal choice of $w$ is $w\simeq 4\sigma_S$~\cite{cressie}.

\begin{figure}
\centerline{\epsfig
{file=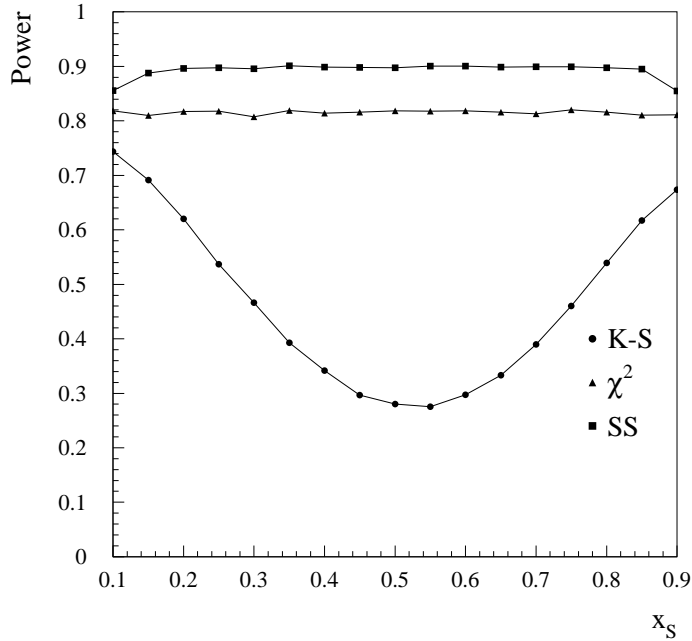,width=10cm} }
\caption{The power of the test statistics versus the peak
positions for $S=20$ and $B\equiv \lambda \Delta =100$.}
\label{fig:compare_narrow} \end{figure}

\begin{figure}
\centerline{\epsfig
{file=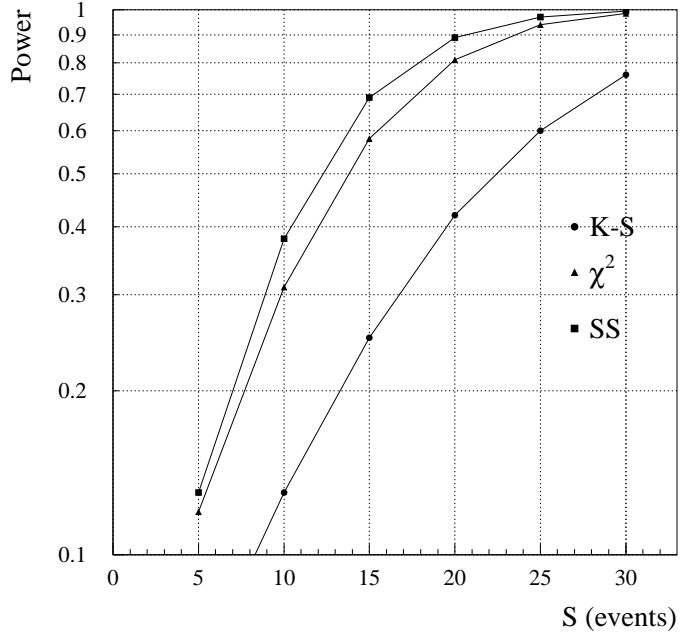,width=10cm} }
\caption{The power of the test statistics averaged over the peak
positions versus the mean expected signal $S$.}
\label{fig:fig_power} \end{figure}

\begin{figure}
\centerline{\epsfig
{file=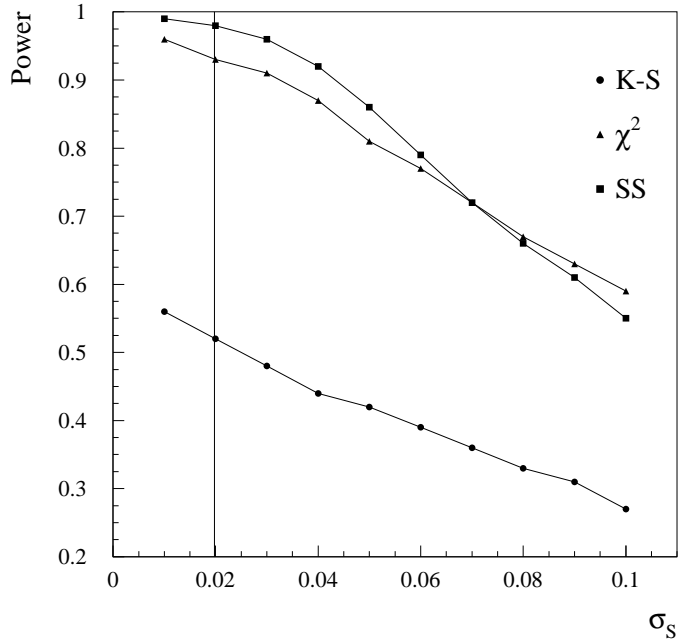,width=10cm} }
\caption{The power of the test statistics averaged over the peak
positions versus the signal width $\sigma_S$ for $B=100$ and
$S=20$. For the Pearson $\chi^2$ test, optimal binning
$N_\mathrm{bin}$ is assumed; for SS
$w=\Delta/N_\mathrm{bin}=0.08$. The vertical line corresponds to
$w=4\sigma_S$.}
\label{fig:compare_broad} \end{figure}

Fig.~\ref{fig:compare_narrow} shows the power of the K-S, SS and
$\chi^2$ tests as a function of the signal position $x_S$. Here,
$\left[\mathcal{A},\mathcal{B}\right]=\left[0,1\right]$,
$\sigma_S=0.05$, $B\equiv \lambda \Delta =100$ and $S=20$.  The
optimal bin size for the $\chi^2$ test has been computed following the
prescription~\cite{moore} $N_\mathrm{{bin}}=2(\lambda \Delta)^{2/5}$,
where $\lambda \Delta$ is the expected sample size in case of null
hypothesis\footnote{Other choices of the binning for the $\chi^2$
test, based on the knowledge of $\sigma_S$, have been tested by Monte
Carlo experimentation. The corresponding powers do not exceed the one
shown in Fig.~\ref{fig:compare_narrow}.}.  In
Fig.~\ref{fig:compare_narrow} signal events generated beyond the
interval $\left[0,1\right]$ are ignored (out of the sensitivity region
$\left[\mathcal{A},\mathcal{B}\right]$~).  The power averaged over the
peak positions is shown in Fig.~\ref{fig:fig_power} as a function of
$S$.  A few comments are in order. As anticipated in
Sec.~\ref{sec:introduction} the K-S test is not appropriate for local
perturbations. The power is limited compared to other statistics and
depends on the peak position, having the highest sensitivity at the
border of the distribution. The Pearson $\chi^2$ test has a much
higher power but in general the peak detection efficiency is reduced
when the peak is shared between two adjacent bins.  On average the
$\chi^2$ test underperforms w.r.t. SS since the correlations among the
bins are ignored\footnote{This is the reason why the $\chi^2$ test and
the Run Test~\cite{eadie} are complementary.}.  However, the bin
prescription for $\chi^2$ is independent of the {\it a priori}
knowledge of $\sigma_S$ while SS makes use of this additional
information. This is a drawback for SS if the cluster width is broader
than the instrumental resolution because the scanning window is no
more optimized. Fig.~\ref{fig:compare_broad} shows the average power
versus $\sigma_S$ assuming $w=\Delta/N_\mathrm{{bin}}$ and $S=20$.
The vertical line corresponds to $w=4\sigma_S$. In fact, it is
possible to compute the scan statistics for an {\it a posteriori}
choice of $w$~\cite{nargawalla} but, clearly, this additional degree
of freedom implies a deterioration of the power.

\begin{figure}
\centerline{\epsfig
{file=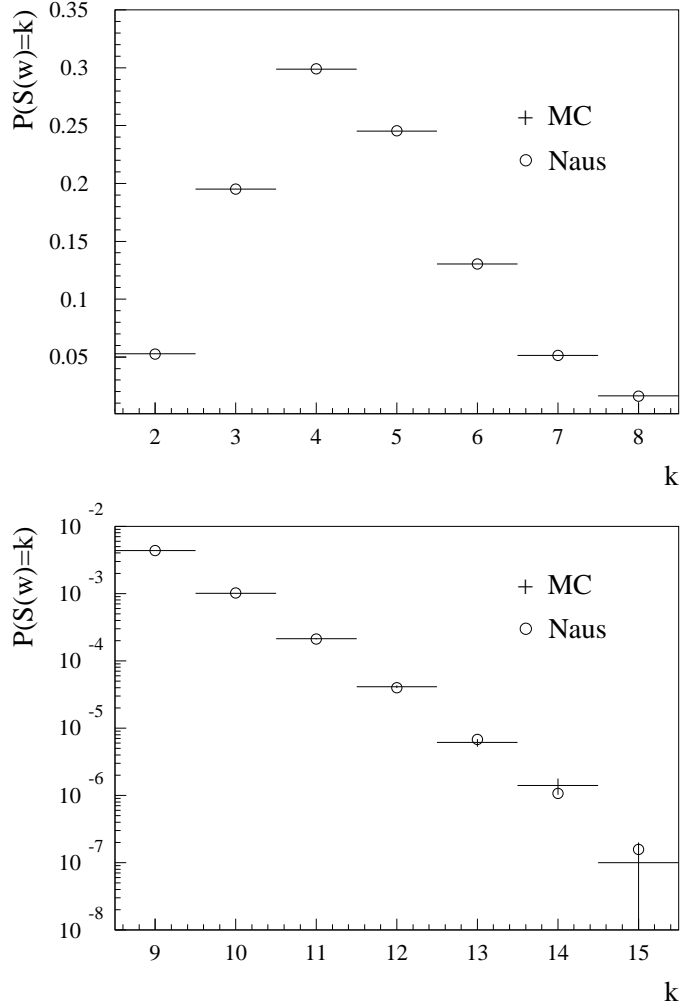,width=10cm} }
\caption{The probability of seeing $k$ events after a scan with
$B=10$, $w=0.2$ under the null hypothesis, computed by
MC (crosses) and Eq.~(\ref{equ:naus}) (empty dots). The high-$k$ tail
of the distribution is shown in the lower plot.}
\label{fig:fig_lowstat} \end{figure}

\begin{figure}
\centerline{\epsfig
{file=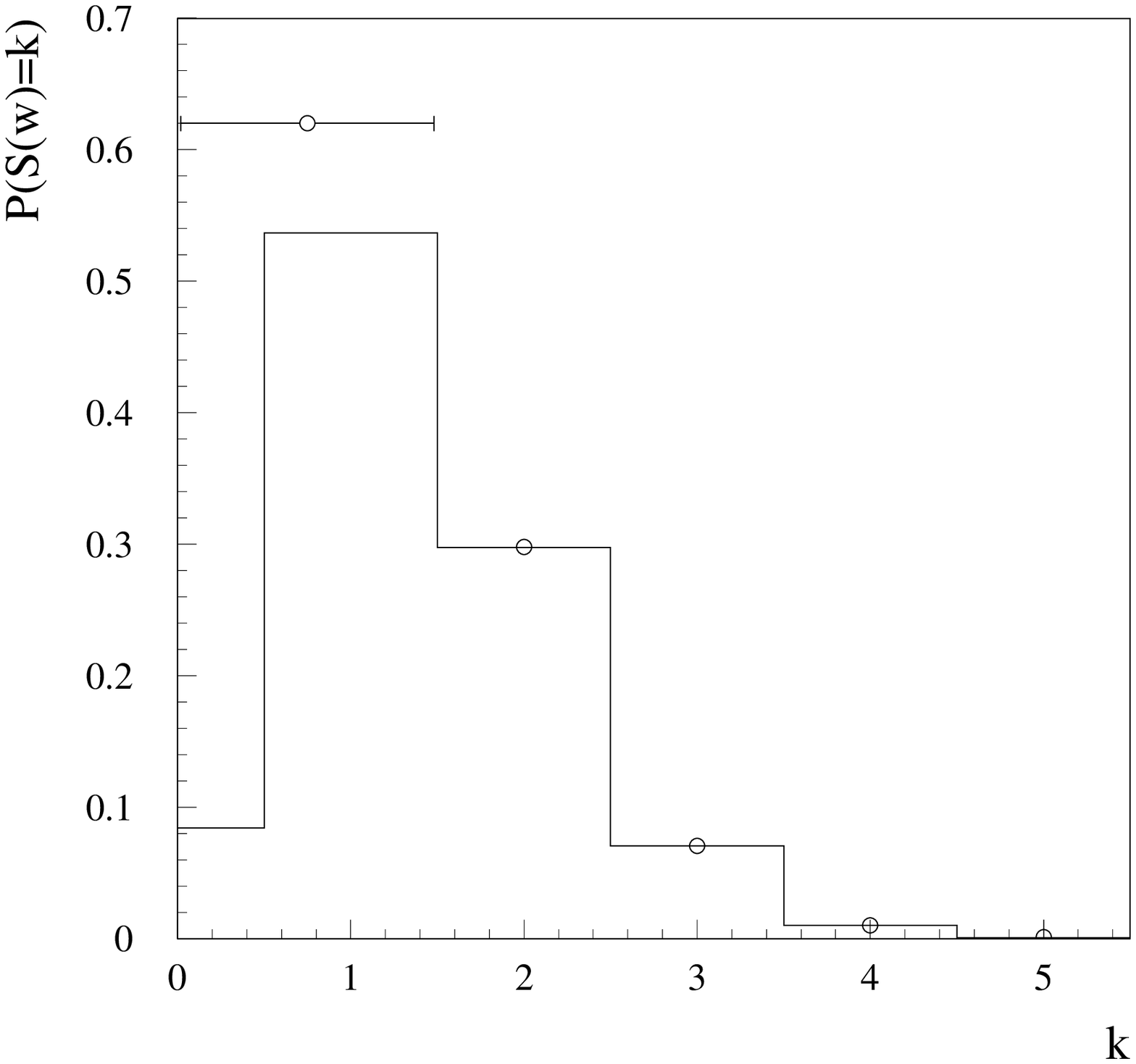,width=10cm} }
\caption{The probability of seeing $k$ events after a scan with
$B=2$, $w=0.2$ under the null hypothesis, computed by
MC (crosses) and Eq.~(\ref{equ:naus}) (empty dots). The first empty
dot indicates the probability of having $k<2$ events after a full
scan: $1-Prob(S(w)=k\geq2)$.}
\label{fig:fig_verylowstat} \end{figure}

On the other hand, SS has a relevant feature which is not manifest in
Figs.~\ref{fig:compare_narrow},\ref{fig:fig_power}. The test
statistics (\ref{equ:test_chi2}) behaves as $\chi^2(N)$ only in the
asymptotic limit.  This implies that the prescription
$N_\mathrm{bin}=2(\lambda \Delta)^{2/5}$ is appropriate only if the
number of expected events per bin is such that the normal limit is
justified. If this is not the case, the extraction of the $p$-value
for the null hypothesis under the assumption $T\sim \chi^2(N)$ is
biased and the proper behavior has to be restored increasing the bin
size or computing the correct $p$-values by MC
experimentation~\cite{heinrich}. This fact is immaterial for SS, since
the derivation of Eqs.~(\ref{equ:exact}) and (\ref{equ:naus}) does not
invoke the Central Limit theorem. The unbiaseness of the $p$-value for
the null hypothesis even for few events expected in the scanning
window has been checked by Monte Carlo experimentation.
Fig.~\ref{fig:fig_lowstat} shows the probability $Prob(S(w)\geq
k)-Prob(S(w)\geq k+1)$ of finding exactly $k$ events after a scan,
computed by Monte Carlo (crosses) and by Eq.~(\ref{equ:naus}). The
upper plot shows the region with highest probability assuming $B=10$,
$S=0$, $w=0.2$; in this case the corresponding $\chi^2$ test with
optimal binning would have no more than 2 events per bin.  The number
of trials is $10^7$ so the MC error in the upper plot is
negligible. The lower plot indicates the tail of the distribution.
Note that the exact formula on which the Naus approximation is based
holds for $k>1$. Biases in the $p$-value will appear only when the
approximation

\be 
Prob(S(w)=0) \simeq 0 \ \ \Longrightarrow 
Prob(S(w)=1) \simeq 1-Prob(S(w) \geq 2) 
\ee

\noindent does not hold, that is when the probability of having zero
events after a full scan is non-negligible as in
Fig.~\ref{fig:fig_verylowstat} where $\lambda \Delta=2$ and the first
empty dot indicates $Prob(S(w)=0 \ \mathrm{or} \ S(w)=1) \ = \
1-Prob(S(w)\geq 2)$.

\section{Extensions of the Scan Statistics}
\label{sec:extensions}

In Sec.~\ref{sec:comparison} the alternative hypotheses (normal
distributions centered at $x_S \in
\left[\mathcal{A},\mathcal{B}\right]$) were such that events generated
outside the sensitivity region $\left[\mathcal{A},\mathcal{B}\right]$
have been ignored. This implies a loss of power for the SS in the case
of $x_S$ lying near the border of the sensitivity region (see
Fig.~\ref{fig:compare_narrow}), i.e. when $x_S-\mathcal{A}$ or
$\mathcal{B}-x_S$ is comparable or smaller than the scanning window
$w$.  Clearly, these results can be extended in a straightforward
manner to a bounded variable $x$ on a range
$\left[\mathcal{A},\mathcal{B}\right]$ where the signal accumulates at
the border. In this case the probability of finding a signal event
between $x$ and $x+\d x$ is

\be
Prob(\hat{x} \in \left[x,x+\d x\right]) = 
\left\{ \begin{array}{r@{\quad:\quad}l}
I^{(-\infty,\mathcal{A})} \ \delta(x-\mathcal{A}) & x=\mathcal{A} \\
G(x,x_S,\sigma_S) & x\in \left(\mathcal{A},\mathcal{B}\right) \\
I^{(\mathcal{B},\infty)} \ \delta(\mathcal{B}-x) & x=\mathcal{B} 
\end{array} \right. 
\label{equ:gau_bounded}
\ee

\noindent and zero for $x$ outside $\left[\mathcal{A},\mathcal{B}\right]$;
$G(x,x_S,\sigma_S)$ is the normal distribution with mean $x_S$ and
variance $\sigma_S^2$ and

\be
I^{(a,b)} \ \equiv \ \int_a^b G(x,x_S,\sigma_S) \d x
\ee

\noindent On the other hand, in many physics applications the
alternative hypothesis~(\ref{equ:gau_bounded}) does not describe our
signal expectation.  For instance, if the bounded variable is
connected with an angular distribution, a signal excess will manifest
as a local perturbation of an uniform distribution of events along a
unit circle. Event positions are described by the bounded variable
$\theta \in \left[0,1\right]$. In this case $S_c(w)$ is defined as the
maximum number of points in any arc of length $w$ and, following the
notation of Sec.~\ref{sec:scanstat} with $\Delta=1$, we
have~\cite{naus82}:

\be
Prob(S_c(w) \geq k) \ \simeq \ 1-Q^*(k,4\psi,1/4)\frac{
\left[Q^*(k,3\psi,1/3)\right]^{L-2} }
{\left[Q^*(k,2\psi,1/2)\right]^{L-1} }
\ee

\noindent 
where $Q^*(k,3\psi,1/3)$ and $Q^*(k,2\psi,1/2)$ are given by 
Eqs.~(\ref{equ:Q2_naus}) and (\ref{equ:Q3_naus}) and $Q^*(k,4\psi,1/4)$ 
can be derived from Eq.~(\ref{equ:exact}) with $L=4$.

Note also that in Sec.~\ref{sec:comparison} we considered the
parameters describing the background known with high precision, so
that it is possible to assume the null hypothesis to be fully
specified. This is not the case if the parameters $\underline{\theta}$
of Eq.~(\ref{equ:b_ab}) have to be estimated {\it after} the data
taking. In this case SS should be extended to devise the optimal
estimate of the underlying background density
$B(x,\underline{\theta})$ that is unbiased and consistent under both
the null hypothesis and the occurrence of a local excess of width
$\sigma_S$. This problem is still unsolved~\cite{glaz_naus} for a
generic function $B(x,\underline{\theta})$. Unbiased estimators have
been obtained for simple functional dependences as in the case of the
linear regression: for a discussion we refer to~\cite{glaz_naus}.

Finally, it is worth mentioning that, even if we focused on 1-dim
distributions, SS has been extended to multivariate
problems~\cite{loader,alm99,auer}. In particular, 2-dim applications
are quite common e.g. in space analysis of arrival direction data of
high energy cosmic rays, x and $\gamma$-ray bursts~\cite{orford}.  A
description of multivariate unconditional scan statistics can be found
in~\cite{glaz_naus,alm99}.

\section{Conclusions}
\label{sec:conclusions}

In this paper, we considered the conditions under which Scan
Statistics can be implemented to signal a departure from the
underlying probability model that describes the experimental data. In
fact, local perturbations (``bumps'' or ``excesses'' of events)
are better dealt within this framework and, in general, tests based on
$S(w)$ provide a powerful and unbiased alternative to the traditional
techniques related with the $\chi^2$ and Kolmogorov distributions.
This holds in particular if the widths of the resonances are known
{\it a priori}, e.g. when the event distribution is dominated by the
instrumental resolution. Approximate formulas for the computation of
SS in the range of interest for high energy and nuclear physics
applications have been provided. Possible extensions to bounded
variables and multivariate problems were also discussed.

\begin{ack}
I'm greatly indebted with L.~Lyons, F.~Ronga and T.~Tabarelli de Fatis for
useful discussions and careful reading of the manuscript.

\end{ack}



\end{document}